\documentclass[conference,10pt]{IEEEtran}

\usepackage{url}
\usepackage{float}
\usepackage{comment}
\usepackage{color}
\usepackage{graphicx}
\usepackage[font=footnotesize,labelfont=bf]{caption}
\usepackage{subcaption}
\usepackage{amsmath,amsfonts,amssymb}
\usepackage{algorithm}
\usepackage{algorithmic}
\usepackage{multirow}

\usepackage{amsthm}
\usepackage{verbatim, url}

\newcounter{theorem}
\newcounter{def}
\newcounter{mylea}
\newcounter{corollary}
\usepackage{fancyhdr}

\begin{document}

\title{\huge \bf FFTPL: An Analytic Placement Algorithm Using \\ Fast Fourier Transform for Density Equalization}


\author{
\IEEEauthorblockN{Jingwei Lu$^1$, Pengwen Chen$^2$, Chin-Chih Chang$^3$, Lu Sha$^3$, \\
Dennis Jen-Hsin Huang$^3$, Chin-Chi Teng$^3$, Chung-Kuan Cheng$^1$}
\IEEEauthorblockA
{$^1$Department of Computer Science and Engineering, University of California, San Diego \\
$^2$Department of Applied Mathematics, National Chung Hsing University \\
$^3$Cadence Design Systems \\
jlu@cs.ucsd.edu, pengwen@nchu.edu.tw, chinchih@cadence.com, lusha@cadence.com, \\
dhuang@cadence.com, ccteng@cadence.com, ckcheng@ucsd.edu\\}}
\maketitle
\thispagestyle{empty}
\renewcommand{\headrulewidth}{0pt}
\begin{abstract}
\label{sec:abs}
We propose a flat nonlinear placement algorithm FFTPL
using fast Fourier transform for density equalization.
The placement instance is modeled as an electrostatic system 
with the analogy of density cost to the potential energy.
A well-defined Poisson's equation is proposed for 
gradient and cost computation.
Our placer outperforms state-of-the-art placers 
with better solution quality and efficiency.
\end{abstract}


\section{Introduction}
\label{sec:intro}

Placement remains an important role in the 
VLSI physical design automation~\cite{abk} 
with impacts to congestion analysis~\cite{cp00},
clock tree synthesis~\cite{dmst}
and routing~\cite{lmgr}.
Placement quality is usually evaluated by
the total half-perimeter wirelength (HPWL), 
which correlates with timing, power, cost and 
routability.
Traditional placement methods can be generally divided 
into four categories.
{\bf Stochastic} approaches~\cite{timberwolf}
are usually based on simulated annealing. 
Despite good solution quality, the runtime is quite long. 
{\bf Min-cut} approaches~\cite{capo} comprises 
recursive problem partitioning and local optimum solution. 
However, improper partitioning would cause 
unrecoverable quality loss.
{\bf Quadratic} approaches~\cite{simpl,rql,fp3} 
approximate the net length using quadratic 
functions 
which enables gradient-based minimization.
By solving the system, 
cells are dragged away from over-filled regions 
with quadratic wirelength overhead. 
Nonetheless, the modeling accuracy remains a long-term issue.
{\bf Nonlinear} approaches~\cite{mpl6,ntupl3,aplace2} refer
to the algorithms using nonlinear optimization framework. 
Wirelength and density are modeled by smooth mathematical functions 
where gradient can be analytically calculated.
Due to the high computation complexity,
nonlinear approaches usually employ multi-level cell
clustering with quality overhead introduced. 

{\bf In this work}, 
we develop a flat nonlinear
placement algorithm 
which produces better and faster solution. 
The placement instance 
is modeled as an electrostatic system 
which induce one density constraint. 
The electric potential and field are coupled with density 
by a well-defined Poisson's equation, which is numerically 
solved using fast Fourier transform (FFT). 
Our algorithm is validated through experiments 
on the ISPD 2005 benchmark suite.

The remainder of the paper is organized as follows.
In Section~\ref{sec:ana}, 
we review the previous works 
and discuss their existing problems. 
In Section~\ref{sec:new}, 
we propose a new formulation of the density constraint 
with numerical solutions.
In Section~\ref{sec:gp} and~\ref{sec:exp}, 
we discuss and validate our placement algorithm.
We conclude the work in Section~\ref{sec:conc}.

\section{Essential Concepts and Related Works}
\label{sec:ana}

Placement instance is formulated 
as a hyper-graph $G=(V,E,R)$ with 
nodes $V$ nets $E$ and region $R$.
Let $V_m$ and $V_f$ denote 
movable nodes (cells) and 
fixed nodes (macros) with $|V_m|=m$.
A placer determines all the cell locations 
$\vec{v}=\left(\vec{x},\vec{y}\right)$, 
where $\vec{x}=(x_1,x_2,\ldots,x_m)$ and 
$\vec{y}=(y_1,y_2,\ldots,y_m)$ are the 
horizontal and veritical cell coordinates.
$\vec{v}$ is named as a placement solution.
We have the placement region $R$ uniformly 
decomposed into $n\times n$ 
rectangular grids (bins) 
denoted as $B$.
The HPWL of each net $e$ is denoted as
$W_e(\vec{v})$ while the total HPWL $W(\vec{v})$ 
is the sum of HPWL of all the nets.
{\footnotesize
\begin{equation}
\label{eq:wle}
W(\vec{v})=\sum_{e\in E}W_e(\vec{v})=\sum_{e\in E}\left(\max_{i,j\in e}|x_i-x_j|+\max_{i,j\in e}|y_i-y_j|\right).
\end{equation}}
Analytic global placement targets minimum total HPWL 
subject to the constraint that the ratio of cell area 
to the site area of every bin $b$ 
(denoted as bin density $\rho_b$) 
does not exceed 
the target density $\rho_t$
\begin{equation}
\label{eq:gp}
\min_{\vec{v}} W\left(\vec{v}\right) \text{ s.t. } \rho_b(\vec{v})\le\rho_t\text{, }\forall b\in B.
\end{equation}
As neither the wirelength function $W(\vec{v})$ 
nor the density function $\rho_b(\vec{v})$ is differentiable, 
smoothing techniques are 
developed to improve the optimization quality.

{\bf Wirelength modeling} functions 
can be divided into two categories.
{\bf Log-Sum-Exp (LSE)} 
wirelength model is proposed 
in~\cite{naylor} and widely used in recent academic 
placers~\cite{mpl6,ntupl3,aplace2}.
{\bf Weighted-Average (WA)} function is recently proposed 
in~\cite{wa} with smaller modeling error 
compared to that of LSE, the equation for the horizontal 
wirelength is 
\begin{equation}
\begin{footnotesize}
\label{eq:wa}
\begin{aligned}
\widetilde{W}_e(\vec{x}) = & \left(  \frac{\sum_{i\in e}x_i\exp\left(\frac{x_i}{\gamma}\right)}{\sum_{i\in e}\exp\left(\frac{x_i}{\gamma}\right)} - \frac{\sum_{i\in e}x_i\exp\left(\frac{-x_i}{\gamma}\right)}{\sum_{i\in e}\exp\left(\frac{-x_i}{\gamma}\right)}  \right).
\end{aligned}
\end{footnotesize}
\end{equation}
\cite{wa} shows that 
the function 
is strictly convex and
converges to HPWL as 
the smoothing parameter $\gamma$ 
approaches zero.

{\bf Density modeling} techniques
generally form two categories.
{\bf Local smoothing} 
functions~\cite{naylor} 
replaces the piece-wise linear original 
density function
with a ``bell-shaped'' quadratic function~\cite{ntupl3,aplace2}.
As only local information is involved,
more iterations may be consumed before the solution 
converges. 
{\bf Global smoothing} techniques 
use elliptic PDE and have many applications 
in modern nonlinear placers~\cite{mpl6}. 
Global information incorporation enables large-scale cell motion. 
Helmholtz equation is proposed in~\cite{mpl6} as below
\begin{equation}
\label{eq:helm}
\Delta\psi(x,y)-\epsilon\psi(x,y)=\rho(x,y)\text{, } (x,y)\in R, 
\end{equation}
where $\psi(x,y)$ is the smoothed density distribution. 
A unique solution can be produced when the linear factor $\epsilon>0$.
However, the smoothing effect becomes sensitive.

{\bf Nonlinear global placement} formulates the 
problem as an unconstrained nonlinear optimization.
In~\cite{ntupl3,aplace2} 
the density constraints are relaxed 
using quadratic penalty method
\begin{equation}
\label{eq:obj1}
\min_{\vec{v}}\text{    }\widetilde{W}(\vec{v})+\lambda\sum_{b\in B}\left(\widetilde{\rho}_b(\vec{v})-\rho_t\right)^2.
\end{equation}
The approach in~\cite{mpl6}
assigns all the grid density $\widetilde{\rho}_b$ 
with penalty factor $\lambda_b$.
However, this consumes longer runtime.

\section{Electrostatic System Modeling}
\label{sec:new}

\begin{figure*}
  \centering
  \begin{subfigure}[b]{0.33\textwidth}
    \centering
    \includegraphics[keepaspectratio, width=1.05\textwidth]{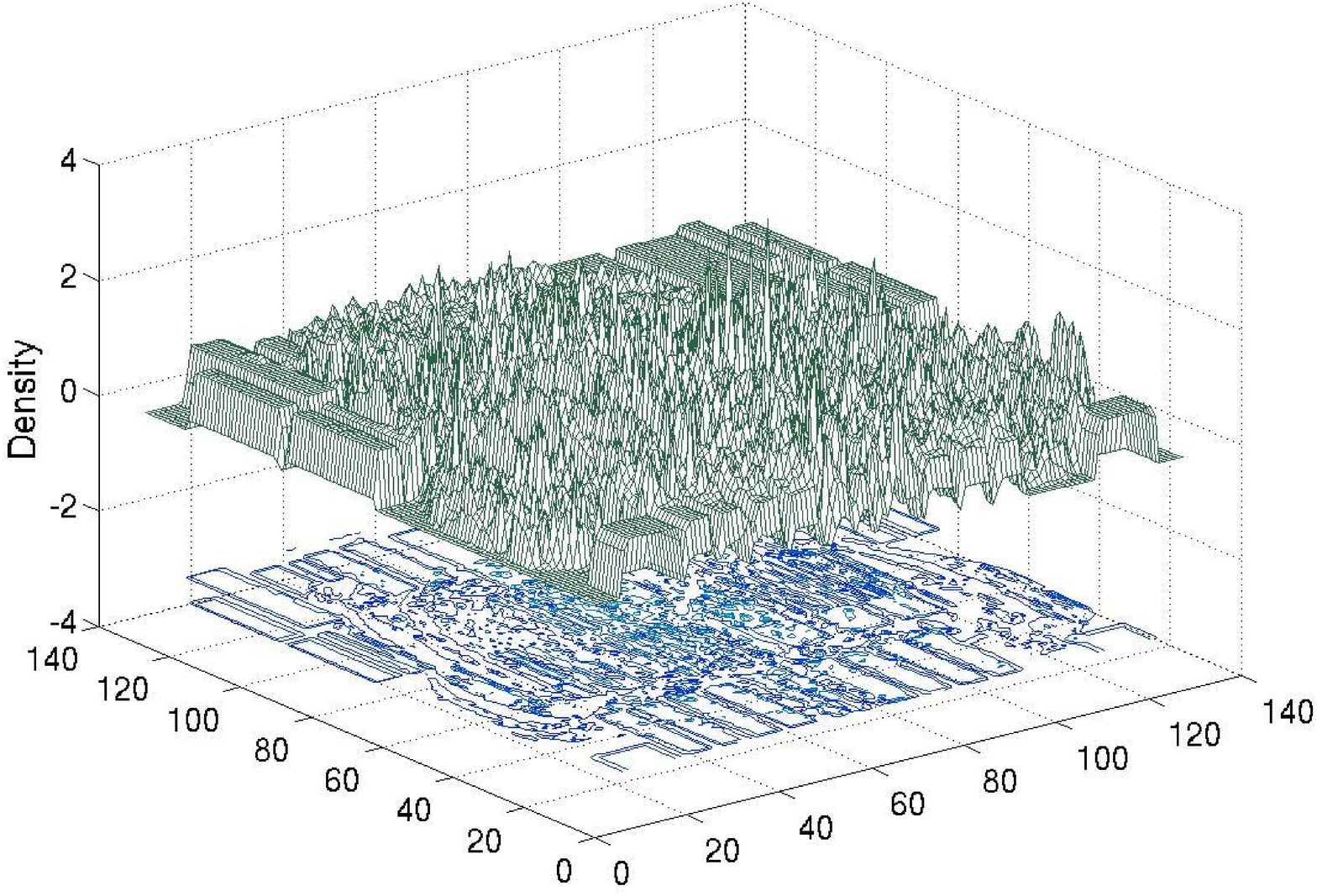}
    \caption{}
    \label{subfig:den}
  \end{subfigure}
  \begin{subfigure}[b]{0.33\textwidth}
    \centering
    \includegraphics[keepaspectratio, width=1.05\textwidth]{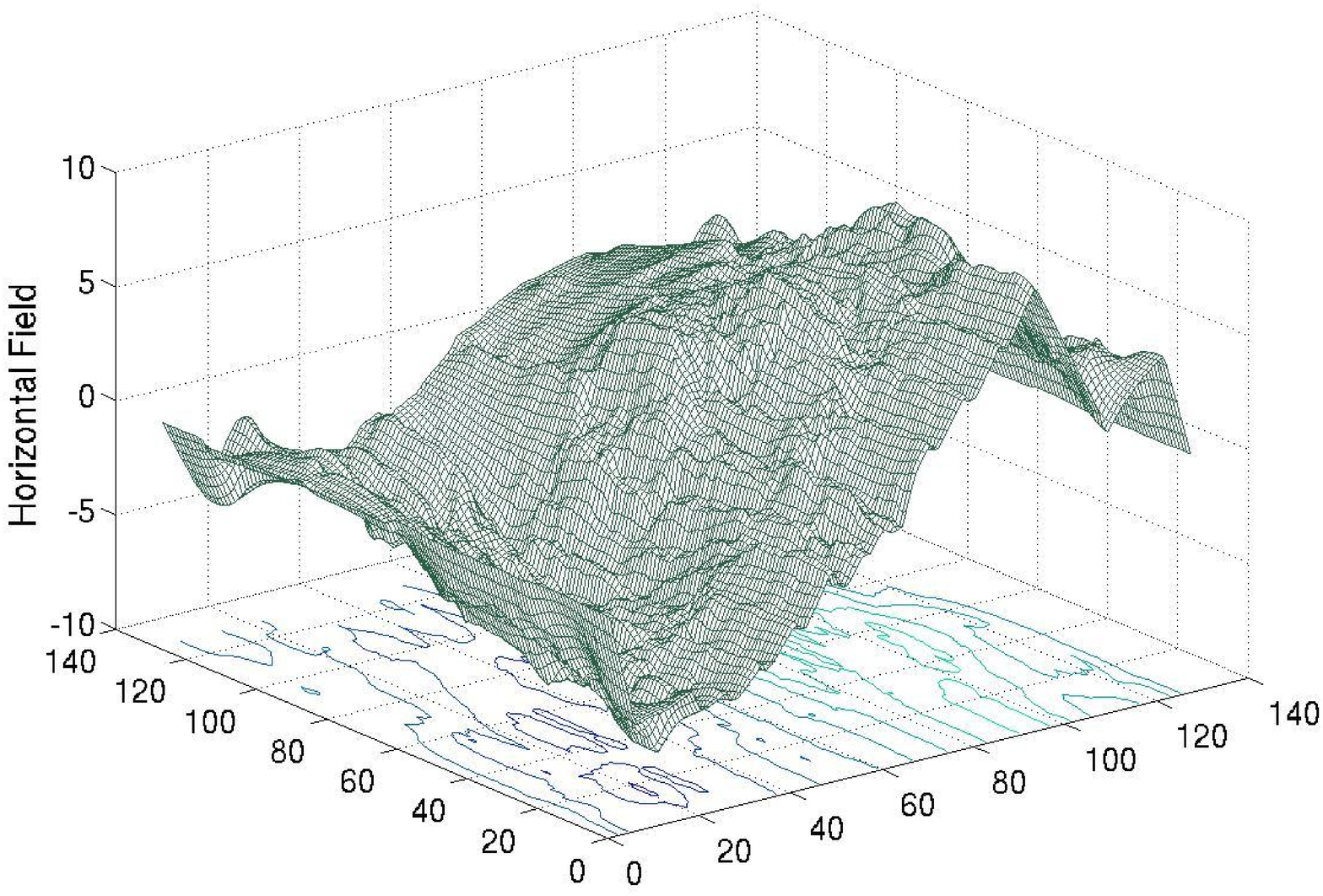}
    \caption{}
    \label{subfig:field}
  \end{subfigure}%
  \begin{subfigure}[b]{0.33\textwidth}
    \centering
    \includegraphics[keepaspectratio, width=1.05\textwidth]{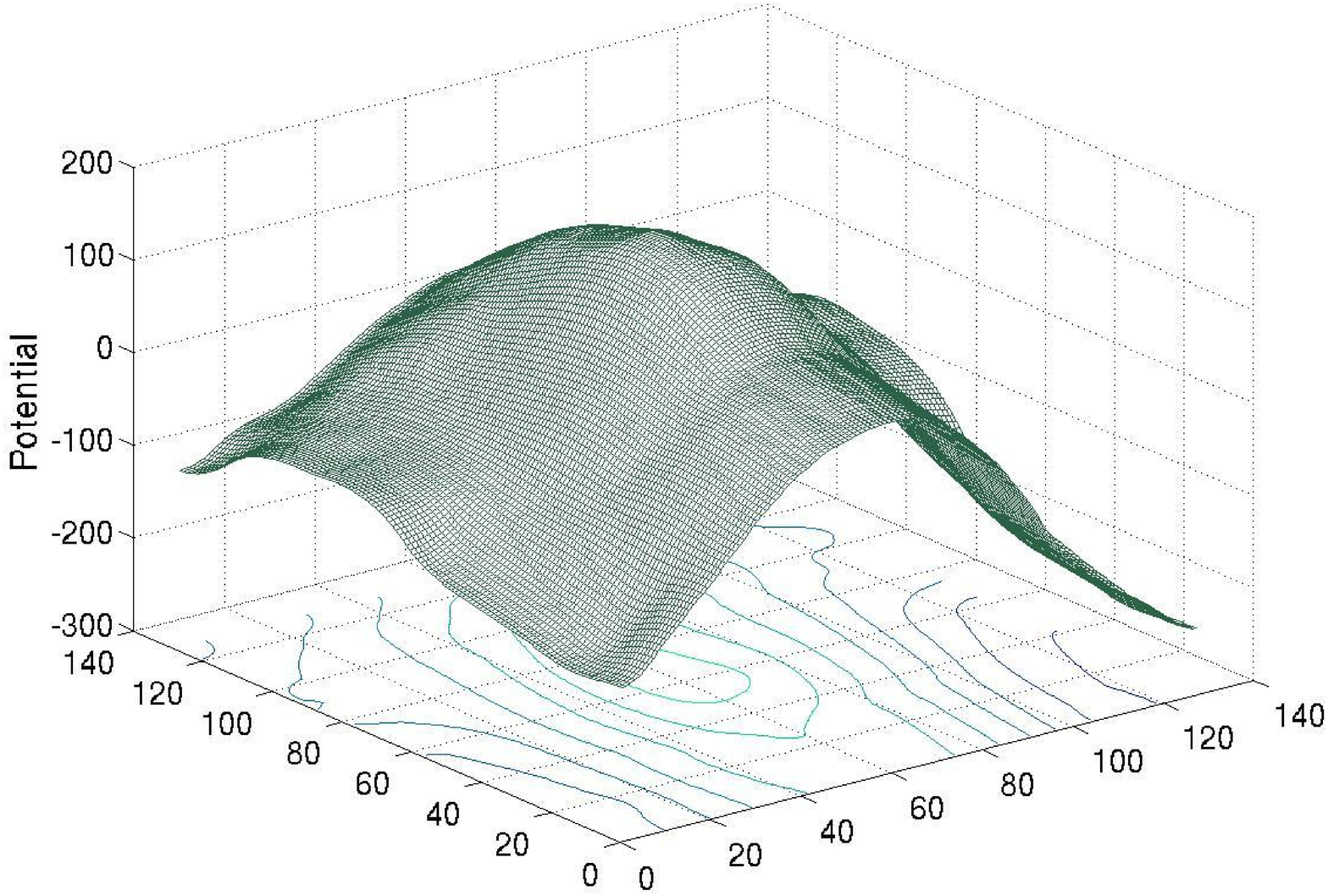}
    \caption{}
    \label{subfig:phi}
  \end{subfigure}
  \caption{The distribution of electric (a) density (b) field (c) potential without filler insertion. Snapshots are extracted at iteration 50 of ADAPTEC1.} 
\label{fig:dfp}
\end{figure*}

We model the placement instance as an 
independent electrostatic system for 
density function transformation. 
Each node $i$ is converted to a
positively charged particle with  
the electric quantity $q_i$ equals
the node area $A_i$. 
Let $\psi_i$ and $E_i$ denote the 
electric potential and field at cell $i$,  
We have the potential energy and electric force 
for each cell as $N_i=q_i\psi_i$ 
and $F_i=q_iE_i$, respectively.
An example is shown in Figure~\ref{fig:dfp}.

\subsection{Electrostatic Equilibrium}

We use electric force for cell movement direction
and density equalization. 
Direct-current (DC) component is removed
from the density function such that 
under-filled regions become negatively charged. 
Cells locating at positive regions are 
attracted for neutralization.
In the end, the system reaches the electrostatic 
equilibrium state with zero bin density and potential 
energy.




\subsection{Potential Energy Computation}

Similar to~\cite{filler,mpl6}, we add disconnected "fillers" to 
induce density force for connected cells clotting thus 
interconnect shortenning.
Placement region could be irregular polygon 
(bounding box $R$). 
We name each non-placeable rectangular region within $R$  
as a ``dark node''.
Cells are pushed away by the density force  
when approaching the chip boundary.
Also, the area of fixed and dark nodes
must be scaled down by the target density 
to globally balance the density force.
Let $V_{fc}$ and $V_d$ denote the sets of 
fillers and dark nodes and 
$V'=V_m\cup V_f\cup V_{fc}\cup V_{d}$, 
the potential energy is computed as
\begin{equation}
\label{eq:potn}
N(\vec{v})=\sum_{i\in V'}N_i=\sum_{i\in V'}q_i\psi_i.
\end{equation}


\subsection{Density Constraint Formulation}

By applying the penalty factor $\lambda$, 
we formulate an unconstrained optimization problem 
\begin{equation}
\label{eq:obj2}
\min_{\vec{v}} f(\vec{v})=\widetilde{W}(\vec{v})+\lambda N(\vec{v}).
\end{equation}
Compared to the quadratic penalty method~\cite{ntupl3,aplace2} or 
the multiple constraints~\cite{mpl6}, 
our method has lower complexity and better quality. 
The gradient vector is 
$\nabla f=\nabla W + \vec{q}\cdot\vec{E}$ where 
$\vec{q}$ and $\vec{E}$ are the electric quantity and 
field vectors of all the cells.
As the electric force always points to the steepest descent 
of system energy, 
we could dynamically balance the wirelength and density forces
using penalty factors.

\subsection{Well-Defined Poisson's Equation}

By Gauss' law, 
the potential and field 
are coupled with density by Poisson's equation.
\begin{equation}
\label{eq:poi}
\begin{cases}
\nabla\cdot\nabla\psi (x,y) = -\rho(x,y)\text{,} \\
\mathbf{\hat{n}}\cdot\nabla\psi (x,y) = \mathbf{0} \text{,     } (x,y)\in \partial R, \\
\iint\limits_R \psi(x,y) = 0\text{.} \\
\end{cases}
\end{equation}
$\hat{n}$ is the outer unit normal and 
$\partial R$ is the boundary of $R$. 
$\nabla\cdot\nabla\equiv\frac{\partial^2}{\partial x^2}+\frac{\partial^2}{\partial y^2}$ is a differential operator.
As electric force decreases to zero at boundary 
to prevent cells from moving outside, 
we select Neumann boundary condition. 
The potential integral is set to zero 
thus Poisson's equation has unique solution.

\subsection{Fast Numerical Solution}

We use fast Fourier transform 
to solve the Poisson's equation~\cite{fft_poi}.
Discrete sine transform (DST) is used  
to represent the electric field, 
which well satisfies the Neumann boundary condition. 
Therefore, potential and density functions are 
represented by discrete cosine transform (DCT).
We first mirror the function domain
from $[0,n-1]\times [0,n-1]$ to 
$[-n,n-1]\times [-n,n-1]$, 
then periodically extend it to infinity.
The density function can thus be expressed as 
\begin{equation}
\label{eq:cos}
\rho(x,y) = \sum_{u}\sum_{v} a_{u,v}\cos(w_ux)\cos(w_vy)\text{,} \\
\end{equation}
where $w_u=\pi\frac{u}{n}$ and 
$w_v=\pi\frac{v}{n}$ are frequency components 
and $a_{u,v}$ are coefficients. 
\begin{equation}
\label{eq:coef}
a_{u,v} = \frac{1}{2n}\sum_{x}\sum_{y} \rho(x,y)\cos(w_ux)\cos(w_vy)\text{.} \\
\end{equation}
The solution to the potential function can be expressed as 
\begin{equation}
\label{eq:psi}
\begin{aligned}
\psi(x,y) = & \sum_{u}\sum_{v}\frac{a_{u,v}}{w_u^2+w_v^2}\cos(w_ux)\cos(w_vy).  \\
\end{aligned}
\end{equation}
Therefore, we have the electric field distribution 
$E(x,y)=(E_x,E_y)$ shown as below
\begin{equation}
\label{eq:dct}
\begin{aligned}
\begin{cases}
E_x (x,y) = \sum_{u}\sum_{v}\frac{a_{u,v}w_u}{w_u^2+w_v^2}\sin(w_ux)\cos(w_vy), \\
E_y (x,y) = \sum_{u}\sum_{v}\frac{a_{u,v}w_v}{w_u^2+w_v^2}\cos(w_ux)\sin(w_vy). \\
\end{cases}
\end{aligned}
\end{equation}
The above equations can be efficiently solved 
using many FFT algorithms~\cite{fft}.
Suppose we have $m$ cells in the netlist.
In each iteration,
we reset the grid density using $O(n^2)$ time 
followed by density update using $O(m)$ time due to netlist traversal.
The FFT computation consumes $O(n^2\log n^2)$ time. 
In Section~\ref{subsec:gp_par} we define $n=O(\sqrt{m})$ thus the 
total complexity is essentially $O(m\log m)$.

\section{Global Placement Algorithm}
\label{sec:gp}
\vspace{-0.05in}

\begin{figure}[http]
\centering
\includegraphics[width=1.0\columnwidth, angle=0]{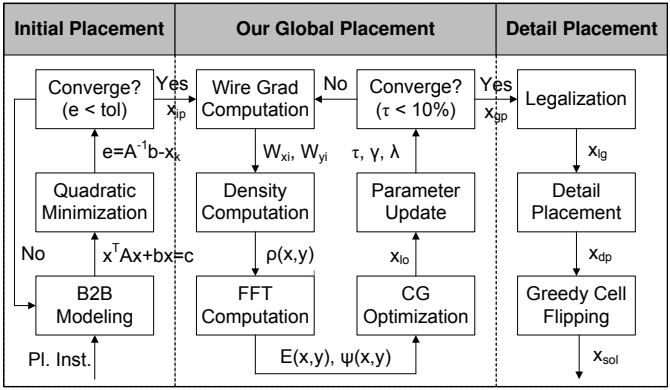}
\caption{The entire flow of initial, global and detail placement.}
\label{fig:flow01}
\end{figure}


The flow of the entire placement optimization is shown 
in Figure~\ref{fig:flow01}.
The initial placement solution $\vec{x}_{ip}$ is based on 
quadratic wirelength minimization 
using bound-2-bound (B2B) net model~\cite{kw2}.
The global placement problem is solved 
using nonlinear Conjugate Gradient (CG) method.
After global placement completes, 
all the filler cells are removed from the 
solution $\vec{x}_{gp}$, which is then legalized 
and discretely optimized
using FastDP~\cite{fastdp}.
with greedy flipping~\cite{capo}.

\subsection{Self-Adaptive Parameter Adjustment}
\label{subsec:gp_par}

\begin{figure}
  \centering
  \begin{subfigure}[b]{0.25\textwidth}
    \centering
    \includegraphics[keepaspectratio, width=1.00\textwidth]{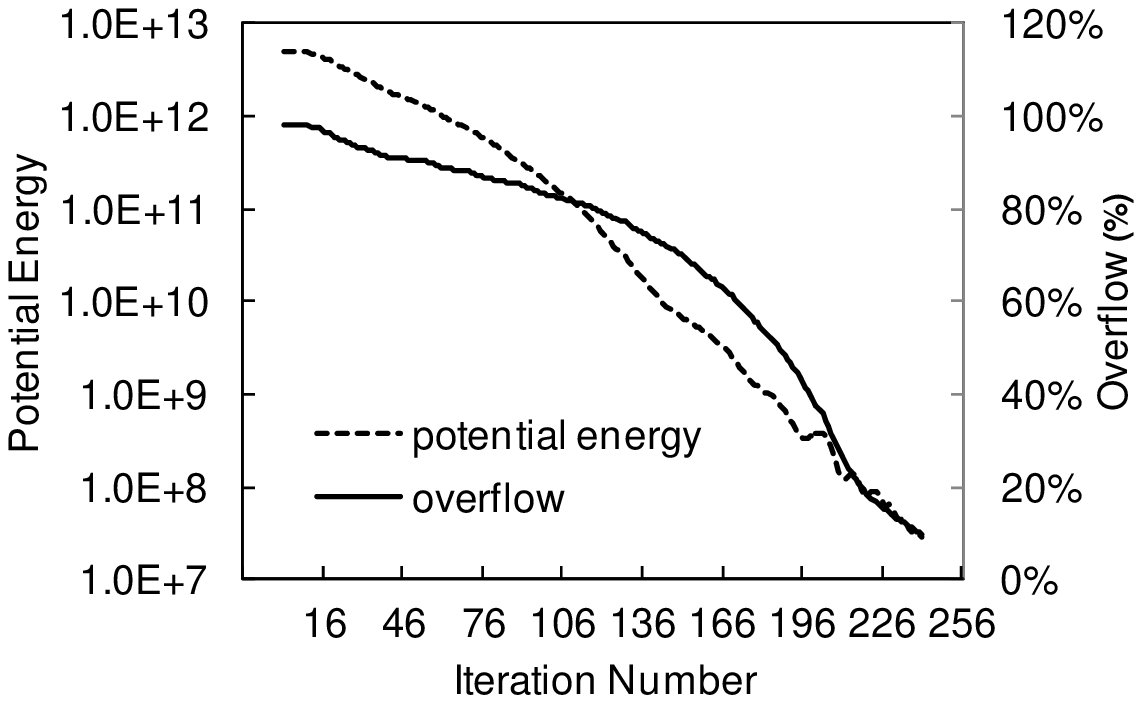}
    \caption{}
    \label{subfig:ovf}
  \end{subfigure}%
  \begin{subfigure}[b]{0.25\textwidth}
    \centering
    \includegraphics[keepaspectratio, width=0.90\textwidth]{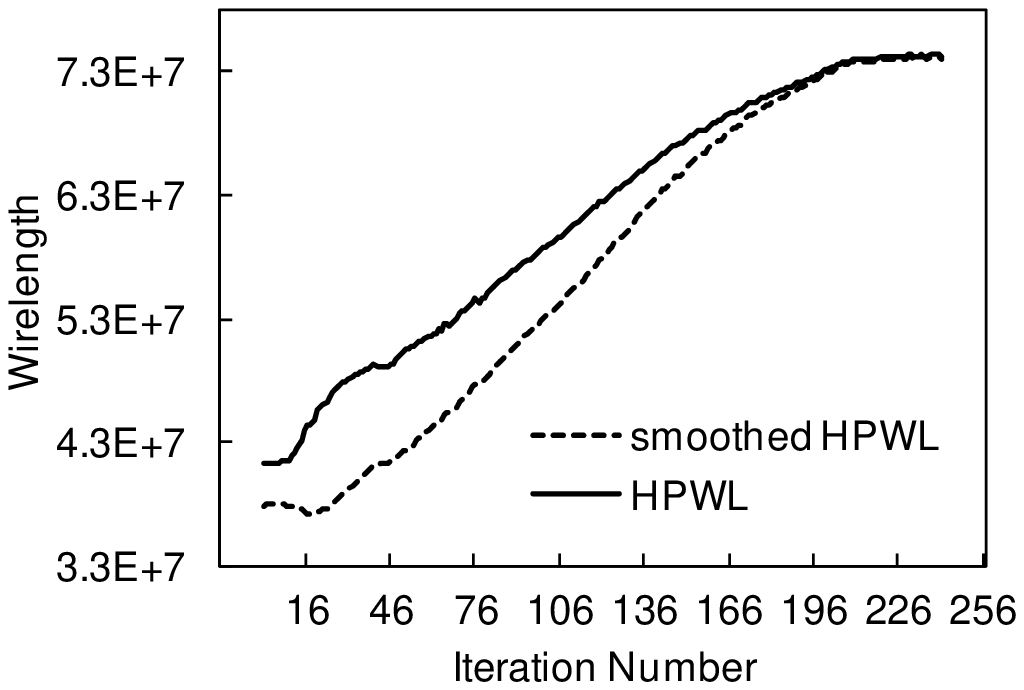}
    \caption{}
    \label{subfig:wlen}
  \end{subfigure}
  \caption{The illustration of (a) total overflow ratio $\tau$ and potential energy $N$ (b) total HPWL $W$ and smoothed wirelength $\widetilde{W}$.}
\label{fig:ovf}
\end{figure}

{\bf Grid dimension} $n$ is statically determined 
before the global placement based on the number of cells 
$m=|A_m|$.
As required to be power of 2 in~\cite{fft}, 
we set $n=\lceil\log_2\sqrt{m}\rceil$ with upper-bound of $1024$.
{\bf Step length} correlates with the search interval of which the 
length is dynamically updated.
The initial value is determined as 
$\alpha^{max}_0=0.044w_b$, where $w_b$ is the grid width.
The search interval is iteratively updated as 
$\alpha^{max}_{k}=\max(\alpha^{max}_0,2\alpha_{k})$ and 
$\alpha^{min}_{k}=0.01\alpha^{max}_k$.
{\bf Penalty factor} is 
initially set as~\cite{ntupl3,aplace2}. 
Unlike those methods with 
constant scaling, 
we iteratively update $\lambda_k=\mu_k\lambda_{k-1}$ 
to balance the wirelength and density forces.
The scaling factor is determined by 
$\mu_k = 1.1^{-\frac{\Delta w_k}{\Delta w_{ref}}+1.0}$
based on HPWL variation 
$\Delta w_k=W(\vec{v}_k)-W(\vec{v}_{k-1})$. 
In practice, we set the reference variation 
$\Delta w_{ref}=3.5\times 10^5$ and bound $\mu_k$ by $[0.75,1.1]$.
{\bf Density overflow} is used to terminate the 
global placement process 
Similar to Eq.~(11) in~\cite{ntupl3},
we use the density overflow $\tau$ as the stopping criterion 
The global placer terminates when $\tau\le 10\%$.
As illustrated in Figure~\ref{fig:ovf}(a), 
system energy is consistent with the density overflow.
{\bf Wirelength coefficient} is used together with WA model~\cite{wa} 
to smooth the HPWL as Figure~\ref{fig:ovf}(b) shows. 
The smoothing parameter $\gamma$ is larger at early time
to encourage global movement 
and smaller at later iterations to enable movement of 
only HPWL-insensitive cells.
We set the smoothing parameter as  
$\gamma = 8.0w_b\times 10^{20/9\times\left(\tau -0.1\right) - 1.0}$.


\subsection{Global Placement}
\label{subsec:gp_flow}

The detail flow of our global placement method FFTPL 
is shown in Algorithm~\ref{alg:gp}. 
We solve the Poisson's equation 
at line 5 by FFT library call~\cite{fft}. 
The global placement solution $x_{gp}$ is output to the legalizer 
and detail placer at line 13. 
\begin{algorithm}
\caption{FFTPL}
\label{alg:gp}
\begin{algorithmic}[1]
\REQUIRE 
initial placement solution $\vec{x}_0=\vec{x}_{ip}$ \\
\ENSURE
global placement solution $\vec{x}_{gp}$
\STATE initialize $\lambda_0$, $\alpha^{max}_0$. 
\FOR {$k=1\to 1000$}
\STATE $f_{k-1}=f(\vec{x}_{k-1})=\widetilde{W}(\vec{x}_{k-1})+\lambda_{k-1} N(\vec{x}_{k-1})$
\STATE compute density function $\rho_{k-1}$
\STATE $(\psi_{k-1},\vec{E}_{k-1})=$FFTsolver$(\rho_{k-1})$
\STATE compute gradient $\nabla f_{k-1}=\nabla\widetilde{W}_{k-1}+\lambda_{k-1}\nabla N_{k-1}$
\STATE $\vec{x}_{k}=$CGsolver$\left(\vec{x}_{k-1},f_{k-1},\nabla f_{k-1},\alpha^{max}_{k-1},0.01\alpha^{max}_{k-1}\right)$
\STATE update $\alpha_k^{max}$, $\lambda_k$, $\tau_k$, $\gamma_k$ 
\IF {$\tau_{k}\le 10\%$} \STATE $\vec{x}_{gp}=\vec{x}_{k}$, break 
\ENDIF
\ENDFOR
\STATE {\bf return} $\vec{x}_{gp}$
\end{algorithmic}
\end{algorithm}


\section{Experiments and Results}
\label{sec:exp}

\begin{table*}[htb]
\caption{HPWL ($\times 10^6$) and runtime (minutes) of all the placers on the ISPD 2005 benchmark suite~\cite{ispd05} 
using official script for performance evaluation.
Experiments are conducted under our 2.67GHz linux machine in single-thread mode. 
Average results are normalized to that of FFTPL (our work).}
\begin{small}
\begin{center}
\begin{tabular}{|c|r|r|r|r|r|r|r|r|r|r|r|r|r|} 
\hline
\multicolumn{2}{|c|}{Categories}   &
\multicolumn{2}{|c|}{Min-Cut}      &
\multicolumn{2}{|c|}{Quadratic}    &
\multicolumn{8}{|c|}{Nonlinear}   \\ \cline{1-14}
\multicolumn{2}{|c|}{Placers}      & 
\multicolumn{2}{|c|}{Capo10.5~\cite{capo}}     &
\multicolumn{2}{|c|}{FastPlace3.0~\cite{fp3}} & 
\multicolumn{2}{|c|}{APlace2~\cite{aplace2}}      &
\multicolumn{2}{|c|}{NTUPlace3~\cite{ntupl3}}    &
\multicolumn{2}{|c|}{mPL6~\cite{mpl6}}         & 
\multicolumn{2}{|c|}{FFTPL}       \\ \hline
Circuits~\cite{ispd05} &\#Cells& HPWL & CPU  & HPWL & CPU  & HPWL & CPU  & HPWL & CPU  & HPWL & CPU & HPWL & CPU \\ \hline
ADAPTEC1   & 211K  & 87.80& 48.33& 78.34& 2.92 & 78.35& 48.88& 80.29& 7.17 & 77.93& 23.27 & 76.46 &  9.17 \\ \hline
ADAPTEC2   & 255K  &102.66& 61.63& 93.47& 4.13 & 95.70& 68.07& 90.18& 8.22 & 92.04& 24.75 & 85.57 & 12.67 \\ \hline
ADAPTEC3   & 452K  &234.27&133.43&213.48& 9.53 &218.52&186.67&233.77& 18.53&214.16& 73.97 & 202.16& 45.40 \\ \hline
ADAPTEC4   & 496K  &204.33&141.85&196.88& 8.75 &209.28&209.60&215.02& 23.53&193.89& 71.03 & 185.83& 34.33 \\ \hline
BIGBLUE1   & 278K  &106.58& 77.90& 96.23& 4.57 &100.02& 64.05& 98.65& 14.30& 96.80& 30.05 & 91.64 & 23.63 \\ \hline
BIGBLUE2   & 558K  &161.68&150.15&154.89& 8.00 &153.75&136.43&158.27& 35.10&152.34& 79.00 & 145.54& 30.83 \\ \hline
BIGBLUE3   & 1097K &403.36&373.87&369.19& 21.05&411.59&289.78&346.33& 38.77&344.10& 104.63& 359.00& 116.67\\ \hline
BIGBLUE4   & 2177K &871.29&730.42&834.04& 40.13&945.77&779.22&829.09&106.08&829.44& 238.82& 805.90& 165.00\\ \hline \hline
\multicolumn{2}{|c|}{Average}  &1.14$\times$  &4.13$\times$  &1.05$\times$  & 0.25$\times$ & 1.10$\times$ & 4.41$\times$ & 1.07$\times$ & 0.66$\times$ & 1.04$\times$ & 1.80$\times$  & 1.00$\times$  & 1.00$\times$  \\ \hline
\end{tabular}
\label{tab:res}
\end{center}
\end{small}
\end{table*}

We implement our algorithm 
using C programming language and 
execute the program in a
Linux operating system with Intel i7 920 2.67GHz CPU
and 12GB memory. 
In our experiments, 
we use the benchmark suite from~\cite{ispd05}.
The target placement density $\rho_t$ is set 
to be $1.0$ for all the benchmarks.
There is no parameter tuning towards specific benchmarks.

We include 
five cutting-edge placers for performance comparison 
with their source code or binary obtained. 
All the results are shown in Table~\ref{tab:res}.
On average, our placer improves the total wirelength by
$13.58\%$,
$5.14\%$,
$10.24\%$,
$7.20\%$ and 
$3.62\%$ 
over
Capo10.5~\cite{capo}, 
FastPlace3.0~\cite{fp3},
APlace2~\cite{aplace2},
NTUPlace3~\cite{ntupl3} and 
mPL6~\cite{mpl6}, 
respectively.
Compared to the published results in 
RQL~\cite{rql} and SimPL~\cite{simpl}, 
our placer produces better solutions  
in six and seven out of the totally eight
benchmarks, respectively.

\section{Conclusion}
\label{sec:conc}
\vspace{-0.05in}

In this paper, 
we propose a flat nonlinear global placement algorithm 
with improved quality and efficiency. 
The placement instance is modeled as an electrostatic system, 
where electric potential and field are computed using Poisson's equation. 
In future, we will extend our algorithm to parallel platform 
and other design objectives (timing, congestion, etc.).

\section{Acknowledgement}
\label{sec:ack}
The authors would like to acknowledge the support of NSF CCF-1017864.

\bibliographystyle{abbrv}
\bibliography{pl}

\begin{thebibliography}{10}

\bibitem{fft}
{General Purpose FFT Package,
  \url{http://www.kurims.kyoto-u.ac.jp/~ooura/fft.html}}.

\bibitem{filler}
S.~N. Adya, I.~L. Markov, and P.~G. Villarrubia.
\newblock {On Whitespace and Stability in Mixed-Size Placement}.
\newblock In {\em ICCAD}, pages 311--318, 2003.

\bibitem{capo}
A.~E. Caldwell, A.~B. Kahng, and I.~L. Markov.
\newblock {Can Recursive Bisection Alone Produce Routable Placements?}
\newblock In {\em DAC}, pages 477--482, 2000.

\bibitem{mpl6}
T.~F. Chan, J.~Cong, J.~R. Shinnerl, K.~Sze, and M.~Xie.
\newblock {mPL6: Enhanced Multilevel Mixed-Size Placement}.
\newblock In {\em ISPD}, pages 212--214, 2006.

\bibitem{ntupl3}
T.-C. Chen, Z.-W. Jiang, T.-C. Hsu, H.-C. Chen, and Y.-W. Chang.
\newblock {NTUPlace3: An Analytical Placer for Large-Scale Mixed-Size Designs
  with Preplaced Blocks and Density Constraint}.
\newblock {\em IEEE TCAD}, 27(7):1228--1240, 2008.

\bibitem{wa}
M.-K. Hsu, Y.-W. Chang, and V.~Balabanov.
\newblock {TSV-Aware Analytical Placement for 3D IC Designs}.
\newblock In {\em DAC}, pages 664--669, 2011.

\bibitem{abk}
A.~B. Kahng, J.~Lienig, I.~L. Markov, and J.~Hu.
\newblock {\em VLSI Physical Design: From Graph Partitioning to Timing
  Closure}.
\newblock Springer, 2010.

\bibitem{aplace2}
A.~B. Kahng, S.~Reda, and Q.~Wang.
\newblock {Architecture and Details of a High Quality, Large-Scale Analytical
  Placer}.
\newblock In {\em ICCAD}, pages 890--897, 2005.

\bibitem{simpl}
M.-C. Kim, D.-J. Lee, and I.~L. Markov.
\newblock {SimPL: An Effective Placement Algorithm}.
\newblock In {\em ICCAD}, pages 649--656, 2010.

\bibitem{dmst}
J.~Lu, W.-K. Chow, C.-W. Sham, and E.~F.~Y. Young.
\newblock {A Dual-MST Approach for Clock Network Synthesis}.
\newblock In {\em ASPDAC}, pages 467--473, 2010.

\bibitem{lmgr}
J.~Lu and C.-W. Sham.
\newblock {LMgr: A Low-Memory Global Router with Dynamic Topology Update and
  Bending-Aware Optimum Path Search}.
\newblock In {\em ISQED}, pages 231--238, 2013.

\bibitem{ispd05}
G.-J. Nam, C.~J. Alpert, P.~Villarrubia, B.~Winter, and M.~Yildiz.
\newblock {The ISPD2005 Placement Contest and Benchmark Suite}.
\newblock In {\em ISPD}, pages 216--220, 2005.

\bibitem{naylor}
W.~C. Naylor, R.~Donelly, and L.~Sha.
\newblock {Non-Linear Optimization System and Method for Wire Length and Delay
  Optimization for an Automatic Electric Circuit Placer}.
\newblock In {\em US Patent 6301693}, 2001.

\bibitem{fastdp}
M.~Pan, N.~Viswanathan, and C.~Chu.
\newblock {An Efficient and Effective Detailed Placement Algorithm}.
\newblock In {\em ICCAD}, pages 48--55, 2005.

\bibitem{timberwolf}
C.~Sechen and A.~Sangiovanni-Vincentelli.
\newblock {TimberWolf3.2: A New Standard Cell Placement and Global Routing
  Package}.
\newblock In {\em DAC}, pages 432--439, 1986.

\bibitem{cp00}
C.-W. Sham, E.~F.~Y. Young, and J.~Lu.
\newblock {Congestion Prediction in Early Stages of Physical Design}.
\newblock {\em ACM TODAES}, 12:1--12, 2009.

\bibitem{fft_poi}
G.~Skollermo.
\newblock {A Fourier Method for the Numerical Solution of Poisson's Equation}.
\newblock {\em Mathematics of Computation}, 29(131):697--711, 1975.

\bibitem{kw2}
P.~Spindler, U.~Schlichtmann, and F.~M. Johannes.
\newblock {Kraftwerk2 - A Fast Force-Directed Quadratic Placement Approach
  Using an Accurate Net Model}.
\newblock {\em IEEE TCAD}, 27(8):1398--1411, 2008.

\bibitem{rql}
N.~Viswanathan, G.-J. Nam, C.~J. Alpert, P.~Villarrubia, H.~Ren, and C.~Chu.
\newblock {RQL: Global Placement via Relaxed Quadratic Spreading and
  Linearization}.
\newblock In {\em DAC}, pages 453--458, 2007.

\bibitem{fp3}
N.~Viswanathan, M.~Pan, and C.~Chu.
\newblock {FastPlace3.0: A Fast Multilevel Quadratic Placement Algorithm with
  Placement Congestion Control}.
\newblock In {\em ASPDAC}, pages 135--140, 2007.

\end{thebibliography}

\end{document}